\documentclass[journal]{IEEEtran}

\ifCLASSINFOpdf
\else
   \usepackage[dvips]{graphicx}
\fi
\usepackage{url}

\usepackage{graphicx, cite}
\usepackage{multirow}
\usepackage{booktabs}
\usepackage{amsmath,amsfonts}
\usepackage[table,xcdraw]{xcolor}
\newcommand{\sunny}[1]{{\color{black}#1}}
\newcommand{\cyc}[1]{{\color{black}#1}}

\begin{document}

\title{An Analysis Method for Metric-Level Switching in Beat Tracking}

\author{Ching-Yu Chiu, Meinard M\"uller,  Matthew E. P. Davies,  Alvin Wen-Yu Su, and Yi-Hsuan Yang
\thanks{Meinard M\"uller is supported by a grant from the International Audio Laboratories Erlangen, which are a joint institution
of the Friedrich-Alexander-Universität Erlangen-Nürnberg (FAU) and Fraunhofer Institute for Integrated Circuits IIS. 
Matthew E. P. Davies is funded by national funds through the FCT - Foundation for Science and Technology, I.P., within the scope of the project CISUC - UID/CEC/00326/2020 and by European Social Fund, through the Regional Operational Program Centro 2020.
}
\thanks{Ching-Yu Chiu is with the Graduate Program of Multimedia Systems and Intelligent Computing, National Cheng Kung University and Academia Sinica, Taiwan (e-mail: sunnycyc@citi.sinica.edu.tw).}
\thanks{Meinard M\"uller is with the International Audio Laboratories Erlangen, Germany (e-mail: meinard.mueller@audiolabs-erlangen.de).}
\thanks{Matthew E. P. Davies is with the Department of Informatics Engineering, Centre for Informatics and Systems of the University of Coimbra, University of Coimbra, Portugal (e-mail: mepdavies@dei.uc.pt).}
\thanks{Alvin Wen-Yu Su is with the Department of CSIE, National Cheng Kung University, Taiwan (e-mail: alvinsu@mail.ncku.edu.tw).}
\thanks{Yi-Hsuan Yang is with Taiwan AI Labs. (e-mail:~yhyang@ailabs.tw).}}

\markboth{Journal of \LaTeX\ Class Files, Vol. 14, No. 8, August 2015}
{Shell \MakeLowercase{\textit{et al.}}: Bare Demo of IEEEtran.cls for IEEE Journals}
\maketitle

\begin{abstract}
For expressive music, the tempo may change over time, posing challenges to tracking the beats by an automatic model. The model may first tap to the correct tempo, but then may fail to adapt to a tempo change, or switch between several incorrect but perceptually plausible ones (e.g., half- or double-tempo). Existing evaluation metrics for beat tracking do not reflect such \sunny{behaviors}, as they typically assume a fixed relationship between the reference beats and estimated beats. In this paper, we propose a new performance analysis method, called annotation coverage ratio (ACR), that accounts for a variety of possible metric-level switching behaviors of beat trackers. The idea is to derive sequences of modified reference beats of all metrical levels for every two consecutive reference beats, and compare every sequence of modified reference beats to the subsequences of estimated beats. We show via experiments on three datasets of different genres the \cyc{usefulness} of ACR \cyc{when being utilized alongside} existing metrics, and discuss the new insights that can be gained. 


\end{abstract}

\begin{IEEEkeywords}
Beat tracking,  evaluation metrics
\end{IEEEkeywords}

\IEEEpeerreviewmaketitle

\section{Introduction}
\label{sec:intro}

\begin{figure}[t]
    \centering
	\includegraphics[width = 0.8\columnwidth]{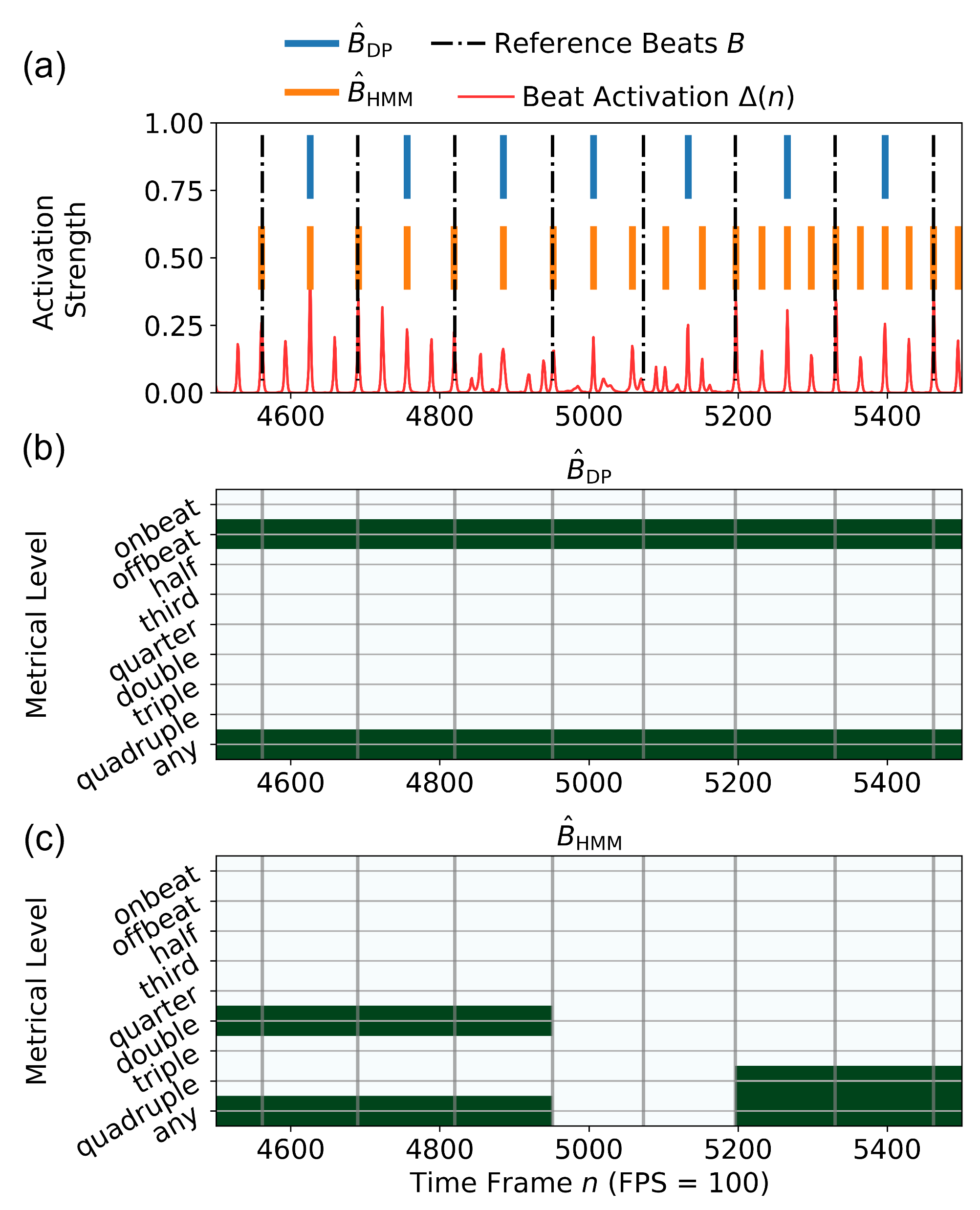}
	\caption{ 
	Visualization of the ACR ($L=2$) evaluation showing the Boolean values whether a reference beat is ``L-correct detected" by the estimated beats w.r.t. different metrical levels. (a) Input beat activation function $\Delta(n)$, reference beats $B$ and the beat estimation of two trackers (DP HMM). \sunny{Visualization of (b) the DP-based and (c) the HMM-based result.}
	}
	\label{fig:acr_vis}
\end{figure}

\IEEEPARstart{B}{eat} tracking 
aims to automatically determine a sequence of time positions that
a listener would tap to when listening to a piece of music \cite{goto94mm,Dixon2000,DP2007}. 
\sunny{Most approaches, which have been found to work well for tracking the beats of music with steady tempo \cite{Bock2016d,Fuentes2018,tcn2019,Bock2020,heydari2021dlb,spl21chiu,9747048}, follow a two-block architecture:}
use a deep learning-based block to generate continuous-valued beat activation functions  $\Delta(n)$ indicating the likelihood of observing a beat at a time instance $n$, and a post processing tracker (PPT) that determines the final beat positions from $\Delta(n)$ using, for example, dynamic programming (DP) \cite{DP2007, fmp2021} or a hidden Markov model (HMM) \cite{Krebs2015, Bock2016mm}. \sunny{See Fig. \ref{fig:acr_vis}(a).}

To quantify the performance of a beat tracker and to study the avenues for improvement, an objective evaluation metric that compares the estimated beat positions $\hat{B} = \{\hat{b}_1, \hat{b}_2, ...\}$
with the reference (ground truth) ones $B = \{b_1, b_2, ...\}$ is needed. 
As the focus has been mostly on tracking music with steady tempo, most existing metrics assume that the relationship between $B$ and $\hat{B}$ with regard to their metrical level is fixed \cite{hainsworth,Klapuri2006,Davies2009}.
For example, $B$ and $\hat{B}$ might be tapped at the same tempo, or $\hat{B}$ might be tapped at a different yet plausible tempo (e.g., half or double tempo) throughout a piece.
Accordingly, if a beat tracker switches its metrical level in the middle of a piece, such metrics may underestimate the beat tracker's performance.
For example, the HMM tracker in Fig. \ref{fig:acr_vis}(a) taps at double tempo first but 
switches to quadruple tempo later; 
existing metrics would only consider its result 
\sunny{to be mostly incorrect}, while perceptually a listener may regard it as performing well most of the time \cite{mckinney2006, mckinney2007, cano2020, cano2021, schreiber2020tempo, Martin2021}.
Such a metric-level switching (MLS) issue has been identified by researchers before \cite{context_dep2007,mckinney2006, pinto2021}, 
\sunny{but there is relatively little work on explicitly investigating MLS.}\footnote{\cyc{Strictly speaking, ``metric-level'' considers only different tempi.  However, as tapping on offbeats (i.e., phases) can be a common mistake (or behavior) of beat trackers, we follow \cite{Davies2009, tempobeatdownbeat:book} and consider a broader definition (with slight abuse of language) that includes offbeats as a type of MLS.}} 


According to our previous attempts to build beat trackers for expressive music that features rich tempo variations \cite{taslp22chiu},  such an MLS issue is prevalent, making it hard to adequately compare the performance of different approaches.  This motivates us to propose a new metric.


\begin{table}[t]
\caption{Conditions considered by different metrics; `($\surd$)' denotes `partially' (cf. Section \ref{sec:reference-variant})}
\setlength{\tabcolsep}{3pt}
\centering
\renewcommand{\arraystretch}{0.8}
\begin{tabular}{ll|cccc}
\toprule
\multicolumn{2}{l}{\multirow{2}{*}{\textbf{Conditions}}} & \multicolumn{4}{c}{\textbf{Evaluation metrics}} \\ 
\multicolumn{2}{c}{}                                       & ~F1   & CMLt/AMLt   & L-correct  & ACR  \\ \midrule
\multicolumn{2}{l|}{\emph{onbeat}}                                 &$\surd$   &$\surd$          &$\surd$         &$\surd$   \\ \hline

\multirow{2}{*}{\emph{offbeat}}     & (half)                    &      &$\surd$          &   $\surd$         &$\surd$   \\ \cline{2-6}
                                 & (one-third,~two-third)                 &      &             &   $\surd$         &$\surd$   \\ \hline
\multirow{2}{*}{\emph{subharmonic}}     & (half,~one-third)                    &      &$\surd$          &            &$\surd$   \\ \cline{2-6}
                                 & (quarter)                 &      &             &            &$\surd$   \\ \hline
\multirow{2}{*}{\emph{harmonic}}        & (double,~triple)                  &      &$\surd$          &            &$\surd$   \\ \cline{2-6}
                                 & (quadruple)               &      &             &            &$\surd$   \\\hline
\multicolumn{2}{l|}{\emph{metric-level switching} (MLS)} &      &             &     ($\surd$)    &$\surd$  \\ \bottomrule
\end{tabular}
\label{tab:conditions}
\end{table}

In this paper, we propose a novel method called ``annotation coverage ratio'' (ACR) that addresses MLS
and considers 
various common metrical levels while evaluating the correctness of each 
estimated beat. 
As listed in Table \ref{tab:conditions},
ACR considers 
a wider variety of metric-level relationships between $B$ and $\hat{B}$ than existing metrics  \cite{hainsworth,Klapuri2006,Davies2009}. 
Moreover, ACR provides a visual tool that facilitates error analysis.
For instance, Fig. \ref{fig:acr_vis}c reveals the MLS behavior of the HMM tracker.

We report experiments on datasets of Western classical, jazz and rock music, showing that ACR leads to new insights and a clearer picture of the performance of beat trackers than existing metrics.
For reproducibility, we open source our code at \sunny{\texttt{\url{https://github.com/SunnyCYC/acr4mls}}}.


\section{Existing Evaluation Metrics}
In this section, we review \sunny{the} commonly used metrics.

\subsection{F1-Score and Continuity-based Evaluation Metrics}
\label{sec:f1}

Given a reference beat sequence $B$ and an estimated one $\hat{B}$,
the F1-score 
\cite{Davies2014,Jia2019,Bock2020,oyama2021}
considers an  
estimated beat $\hat{b}_j \in \hat{B}$ as correct, if 
there is a reference beat $b_i \in B$ that falls within a tolerance window $\varepsilon$ (e.g., $\pm 70$ms) to $\hat{b}_j$: $|b_i-\hat{b}_j| \leq \varepsilon$. 
Specifically, for a test set, we first calculate 
the precision $\mathrm{P}$ (the proportion of estimated beats that are considered correct) and the recall $\mathrm{R}$ (the proportion of reference beats that are found), and then compute $\mathrm{F1=2PR/(P+R)}$. 


Instead of evaluating each beat individually, \emph{continuity}-based metrics \cite{hainsworth,Klapuri2006,Davies2009, tempobeatdownbeat:book} further take into account the inter-beat interval (IBI), i.e., the temporal difference between two adjacent beats.
For $\hat{b}_j$ to be considered as correct, not only both $\hat{b}_j$ and $\hat{b}_{j-1}$ have to fall within the tolerance window around their corresponding reference beats (i.e., $b_i, b_{i-1}$), but also 
their IBIs have to be similar up to a factor of $\gamma$ (i.e.,
$|(b_i-b_{i-1})-(\hat{b}_j-\hat{b}_{j-1})| \leq \gamma (b_i-b_{i-1})$). Commonly, in the literature, the factor $\gamma=0.175$ is used.
There are two main variants \cite{Davies2009}:
CMLt (correct metrical level) only compares $B$ with $\hat{B}$, while AMLt (allowed metrical levels) synthetically creates variants of $B$ that tap at a different yet plausible metrical level (e.g., half offbeat), compares each of the variant with $\hat{B}$, and outputs the best score. See Section \ref{sec:reference-variant} for 
details.

\subsection{L-Correct Detection}
\label{sec:l-correct}
Grosche \emph{et al.}\cite{Grosche2011} proposed a context-sensitive evaluation method, called ``L-correct detection,'' where the parameter $L\in \mathbb{N}_{\geq 2}$ specifies the length of the temporal context in beats. The  idea is to consider every reference beat $b_i$ as an \emph{instance} and check \emph{instance by instance} (e.g., from $i=1$ onwards) if the subsequence 
$B_{i,L}=\{b_i, b_{i+1}, ..., b_{i+L-1}\}$ containing $L$ consecutive reference beats starting from $i$ can be fully matched by \sunny{some} subsequence of the estimated beats $\hat{B}_{j,L} = \{\hat{b}_j, \hat{b}_{j+1}, ..., \hat{b}_{j+L-1}\}$ that starts from $j$. 
This implies $|b_{i+\tau}-\hat{b}_{j+\tau}| \leq \varepsilon$ for $\tau \in \{0,1,\dots,L-1\}$, not allowing any false-positives or false-negatives.
In case of a match,
all the beats in $B_{i,L}$ are considered \emph{L-correct detected}.
This ensures each L-correct detected reference beat to be within the group of at least $L$ consecutive estimated beats. 
Using the number of \emph{L-correct detected} beats, Grosche \emph{et al.}\cite{Grosche2011} \sunny{defined the L-correct recall, precision, and 
the 
F-measure $\mathrm{F}^{L}$.}

\subsection{Creation and Usage of Variants of Reference Beats}
\label{sec:reference-variant}


As shown in Table \ref{tab:conditions}, both AMLt and L-correct consider further the \emph{half offbeat} variants of $B$, taking the middle of every two adjacent beats from $B$ to create $\beta_i \equiv (b_i+b_{i+1})/2$.
AMLt uses a new sequence $B^\text{off-1/2} = (\beta_1, \beta_2, \dots)$ 
to compare with the whole sequence of $\hat{B}$, hence allowing 
a beat tracker to tap at half offbeats. 
However, according to whether $\hat{B}$ matches better to $B$ or to $B^\text{off-1/2}$, AMLt reports only the higher matching score. Namely, it assumes the beat tracker to tap either at onbeats, or at half offbeats, \emph{throughout the piece}, thereby not allowing for MLS.
In contrast, for the L-correct metric in \cite{Grosche2011}, the evaluation is processed instance by instance. In other words, a beat $b_i$ is considered \sunny{L-}correct if there is a matching \sunny{estimated} sequence either for $B_{i,L}$ (onbeat case) or for $B^\text{off-1/2}_{i,L}$ (offbeat case).
This great flexibility is desirable to address the MLS issue.

AMLt and L-correct have their own strengths and weakensses. 
AMLt cannot deal with MLS but it considers actually not only the offbeat  but also the \emph{subharmonic} and \emph{harmonic} cases (i.e., integer divisions or multiples of a certain tempo, such as half tempo and double tempo \cite{fmpTmpharmonic}; see Section \ref{sec:acr} for details).
Furthermore, the L-correct metric partially addresses the MLS between onbeats and offbeats, but not the other metrical levels. 
The proposed ACR combines the advantages of AMLt and the L-correct metric by considering much more metric-level conditions while evaluating the matching of beats instance by instance.

\begin{figure}[t]
	\includegraphics[width = 0.9\columnwidth]{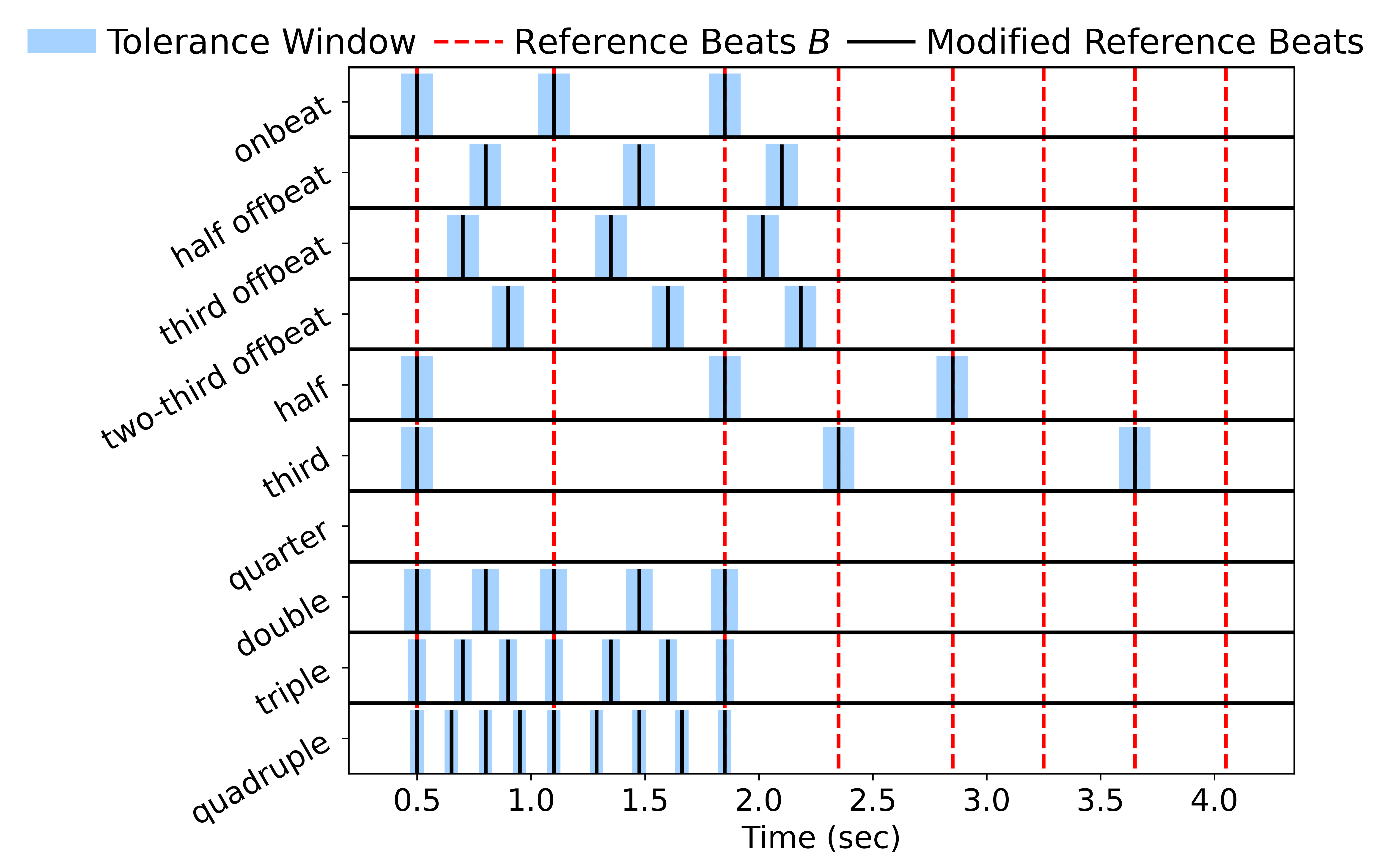}
	\caption{
	Given a sequence of reference beats $\{b_1, b_2, ..., b_8\}$, the figure shows the variants of the subsequence with length $L=3$ created for the first instance for different metrical levels. \cyc{The grid corresponds to the reference beats.}
	}
	\label{fig:refvar_2steps}
\end{figure}

\section{Annotation Coverage Ratio}
\label{sec:acr}


\subsection{Subharmonic and Harmonic Reference Beats}

We describe below how we extend the L-correct metric to account for both tempo subharmonics and harmonics.
Figure \ref{fig:refvar_2steps} visualizes the variants of reference beats adopted by ACR.

To account for of subharmonic tempi, 
we \emph{sub-sample} $B$ from the $i$-th instance onwards and derive the following:
\begin{equation} \label{eq:subharmonic_eq}
\begin{aligned}
&B^\text{half}_{i,L} &= \{b_i, b_{i+2}, \dots, b_{i+2(L-1)}\} \,,\\ 
&B^\text{third}_{i,L} &= \{b_i, b_{i+3}, \dots, b_{i+3(L-1)}\} \,,\\ 
&B^\text{quarter}_{i,L} &= \{b_i, b_{i+4}, \dots, b_{i+4(L-1)}\} \,. 
\end{aligned}
\end{equation}
If the sequence of reference beats $B$ is not long enough to generate a modified reference beats of a specific metrical level `*', we assign $B^*_{i, L}= \emptyset$ (see Figure \ref{fig:refvar_2steps} for $*=\text{quarter}$).

To account for harmonic tempi, we 
\emph{upsample} $B$ 
by linear interpolation, namely by taking
the bisection points (for double tempo), trisection points (for triple tempo), and quadrisection points (for quadruple tempo) for the interval $\overline{b_m b_{m+1}}, m \in[i: i+L-2]$ as additional reference beat positions. Due to the interpolation, the length of the resulting variants will be
$L'=L+(h-1)(L-1)$, with $h=2, 3, 4$ corresponding to the double, triple, and quadruple cases. We therefore consider a longer context $L'$ for L-correct detection of the harmonic cases.
Denoting the $\tau$-th element of the variant created at instance $i$ as
$B^*_{i,L'}[\tau]$, where $\tau$ starts from 0, we say that $B^*_{i,L'}$ is L-correct detected if there exists an index $j$ such that 
$|B^*_{i,L'}[\tau]-\hat{b}_{j+\tau}| \leq \varepsilon$, for $\tau \in \{0,1,\dots,L'-1\}$.


Conventionally, the tolerance window $\varepsilon$ is usually set to a fixed value, such as 70ms. However, due to the upsampling, the IBI for some adjacent beats in $B^*_{i,L'}$ (i.e., $B^*_{i,L'}[\tau+1]-B^*_{i,L'}[\tau]$) might be already smaller than 70ms.
To accommodate this, we instead compute the average IBI of $B^*_{i,L'}$, denoted as $\bar{\text{IBI}}_{i, L'}^*$, and use an adaptive $\varepsilon = \min( \text{70\,ms}, \gamma \, \bar{\text{IBI}}_{i, L'}^*)$, with $\gamma=0.175$ following the convention of CMLt/AMLt.
Figure \ref{fig:refvar_2steps} shows how $\varepsilon$ gets smaller for the tempo harmonic cases. 


For highly expressive music, the additional beat positions created by linear interpolation may not be musically optimal \cite{Desain1993}. We consider our approach mainly as an analysis tool and leave the discussion of better interpolation for future work.

\subsection{Visualization and Annotation Coverage}

For every reference beat position $b_i\in B$, we propose to check if it is \emph{L-correct detected} w.r.t. any of the ten metric-level conditions shown in Figure \ref{fig:refvar_2steps}, and use a separate Boolean value to indicate the result.
This way, we can plot the Boolean values for all the conditions in one graph to visualize whether and how each individual \sunny{beat $b_i$ is L-correct} w.r.t. different conditions. 
As exemplified in Fig. \ref{fig:acr_vis}, as long as $b_i$ is L-correct detected w.r.t. a condition (e.g., offbeat), 
\sunny{it}
will be ``covered'' by a colored bar in that row.
We define the ``annotation coverage ratio'' (ACR) \sunny{of} each condition as the percentage of reference beats that \sunny{are} covered w.r.t. that condition.

To have a more general view of the overall performance, we additionally define an \emph{any tempo} case (shown in the bottom of Fig. \ref{fig:acr_vis}b and \ref{fig:acr_vis}c) that takes the union of the Boolean values of all metric-level cases per beat $b_i$, leading to `ACR-any.'
Similarly, we take the union of the three offbeat cases to derive `ACR-offbeat' (shown in the second row of Fig. \ref{fig:acr_vis}b and \ref{fig:acr_vis}c). Note that, by definition, ACR cannot exceed 1.0 under any condition. \cyc{To monitor and prevent inappropriately high MLS frequency, we define MLS ratio (MLSR) as the proportion of covered reference beats which switch metrical levels.}
\footnote{\cyc{One may also increase $L$ value to achieve stricter metric-level criteria. 
}}

\begin{table}[t]
\caption{Statistics of the datasets.}
\setlength{\tabcolsep}{6pt}
\centering
\begin{tabular}{l|c|c|c|c}
\toprule
\multirow{2}{*}{\textbf{Data}} & \multirow{2}{*}{\textbf{\# tracks}} & \textbf{Total}    & \textbf{\% stable} & \textbf{Mean track}  \\
                               &                                     & \textbf{duration} & \textbf{tempi}     & \textbf{tempo (BPM)} \\ \midrule
Maz-5\cite{Grosche2010b}                          & 301                                 & 12h 27m           & 13.1\%             & 125.39               \\
RWC-Jazz~\cite{rwcdatabase}                       & \,~50                                  & \,~3h 42m            & 74.8\%             & \,~89.67                \\
Rock~\cite{rockdataset}                           & 200                                 & 12h 53m           & 81.4\%             & 115.68              
\\\bottomrule                                      
\end{tabular}
\label{tab:dataset}
\end{table}

\section{Experiment Setup}

In our experiment, we use the three datasets listed in Table \ref{tab:dataset} to investigate the effectiveness of ACR for three important music genres---Western classical, jazz, and rock.
Maz-5 \cite{Grosche2010b} is a private collection of
301 recordings of five
Chopin Mazurkas collected within the Mazurka Project \cite{maz2010} and annotated by Sapp \cite{Sapp2008}.
RWC-Jazz \cyc{(or `Jazz' for short)} contains 50 jazz recordings produced for the RWC Database \cite{rwcdatabase}.
The Rock dataset \cite{rockdataset} contains 200 of the ``500 Greatest Songs of All Time'' listed the Rolling Stone magazine.
Along with other statistics, Table \ref{tab:dataset} also shows the
average tempo per track and the ``percentage of stable tempi'' \cite{Schreiber2020a} of the datasets.\footnote{Following \cite{Schreiber2020a}, after converting all IBIs of each dataset into tempo values and dividing these values with their corresponding mean track tempo, we can derive the normalized tempi and calculate the ``percentage of stable tempi'' as the proportion of normalized tempi that falls in the commonly adopted $\pm 4\%$ tolerance interval of stable tempi \cite{gouyon2006}.}
We see that Maz-5 has the lowest number of stable tempi, while Jazz has the slowest average tempo.

\begin{table*}[t]
\caption{The performance of different post-processing trackers (PPTs) on different datasets using different evaluation metrics. }
\setlength{\tabcolsep}{3pt}
\centering
\begin{tabular}{l|c|c|c|c|c|c|c|c|c|c|c|c|c|c|c|c|c}
\toprule
                                &                                &                               &                              &                              &                                 &                                 &                                             \textbf{L-correct} & \multicolumn{10}{c}{\textbf{ACR (L=2)}}                                                                                            \\ \cline{9-18}
                                &                                &                               &                              &                              &                                 &                                 &                                            &                                &                                   &                                    & \multicolumn{3}{c|}{\textbf{subharmonics}}                                                     & \multicolumn{3}{c|}{\textbf{harmonics}}                                                        &                                 \\ \cline{12-17}
\multirow{-3}{*}{\textbf{Data}} & \multirow{-3}{*}{\textbf{PPT}} & \multirow{-3}{*}{\textbf{F1}} & \multirow{-3}{*}{\textbf{R}} & \multirow{-3}{*}{\textbf{P}} & \multirow{-3}{*}{\textbf{CMLt}} & \multirow{-3}{*}{\textbf{AMLt}} & \multirow{-2}{*}{\textbf{$\mathrm{F}^{L}$($L=2$)}} & \multirow{-2}{*}{\textbf{any}} & \multirow{-2}{*}{\textbf{onbeat}} & \multirow{-2}{*}{\textbf{offbeat}} & \textbf{half}                 & \textbf{third}                & \textbf{quarter}              & \textbf{double}               & \textbf{triple}               & \textbf{quadruple}            & \multirow{-2}{*}{\textbf{MLSR}} \\ \midrule
                                & DP                             & 0.488                         & 0.493                        & 0.468                        & 0.315                           & 0.315                           & 0.590                                      & \cellcolor[HTML]{FFFFFF}0.598  & \cellcolor[HTML]{FFEDB7}0.383     & \cellcolor[HTML]{FFD966}0.181      & \cellcolor[HTML]{FFFFFF}0.025 & \cellcolor[HTML]{FFFFFF}0.001 & \cellcolor[HTML]{FFFFFF}0.000 & \cellcolor[HTML]{FFE697}0.013 & \cellcolor[HTML]{FFD966}0.002 & \cellcolor[HTML]{FFD966}0.001 & \cellcolor[HTML]{FFD966}0.006   \\
                                & HMM                            & 0.499                         & 0.393                        & 0.752                        & 0.098                           & 0.248                           & 0.128                                      & \cellcolor[HTML]{FFF0BF}0.726  & \cellcolor[HTML]{FFFFFF}0.112     & \cellcolor[HTML]{FFFFFC}0.007      & \cellcolor[HTML]{FFD966}0.283 & \cellcolor[HTML]{FFD966}0.235 & \cellcolor[HTML]{FFD966}0.128 & \cellcolor[HTML]{FFFFFF}0.003 & \cellcolor[HTML]{FFF8E1}0.001 & \cellcolor[HTML]{FFFFFF}0.000 & \cellcolor[HTML]{FFF1C4}0.002   \\
\multirow{-3}{*}{Maz-5}         & SPPK                           & 0.822                         & 0.754                        & 0.918                        & 0.569                           & 0.570                           & 0.743                                      & \cellcolor[HTML]{FFD966}0.904  & \cellcolor[HTML]{FFD966}0.682     & \cellcolor[HTML]{FFFFFF}0.003      & \cellcolor[HTML]{FFDB6B}0.276 & \cellcolor[HTML]{FFF4D3}0.070 & \cellcolor[HTML]{FFF8E2}0.025 & \cellcolor[HTML]{FFD966}0.018 & \cellcolor[HTML]{FFFFFF}0.000 & \cellcolor[HTML]{FFFFFF}0.000 & \cellcolor[HTML]{FFE184}0.005   \\ \midrule
                                & DP                             & 0.781                         & 0.796                        & 0.765                        & 0.787                           & 0.845                           & 0.895                                      & \cellcolor[HTML]{FFD966}0.915  & \cellcolor[HTML]{FFD966}0.790     & \cellcolor[HTML]{FFD966}0.124      & \cellcolor[HTML]{FFFFFF}0.000 & \cellcolor[HTML]{FFFFFF}0.000 & \cellcolor[HTML]{FFFFFF}0.000 & \cellcolor[HTML]{FFFFFF}0.001 & \cellcolor[HTML]{FFFFFF}0.000 & \cellcolor[HTML]{FFFFFF}0.000 & \cellcolor[HTML]{FFFFFE}0.000   \\
                                & HMM                            & 0.797                         & 0.889                        & 0.746                        & 0.649                           & 0.803                           & 0.674                                      & \cellcolor[HTML]{FFDD73}0.903  & \cellcolor[HTML]{FFE79D}0.669     & \cellcolor[HTML]{FFFAEA}0.026      & \cellcolor[HTML]{FFFFFE}0.000 & \cellcolor[HTML]{FFFFFF}0.000 & \cellcolor[HTML]{FFFFFF}0.000 & \cellcolor[HTML]{FFEBAD}0.192 & \cellcolor[HTML]{FFFBEF}0.000 & \cellcolor[HTML]{FFDC70}0.016 & \cellcolor[HTML]{FFFFFF}0.000   \\
\multirow{-3}{*}{Jazz}          & SPPK                           & 0.680                         & 0.822                        & 0.601                        & 0.336                           & 0.568                           & 0.434                                      & \cellcolor[HTML]{FFFFFF}0.774  & \cellcolor[HTML]{FFFFFF}0.452     & \cellcolor[HTML]{FFFFFF}0.010      & \cellcolor[HTML]{FFD966}0.029 & \cellcolor[HTML]{FFD966}0.004 & \cellcolor[HTML]{FFD966}0.000 & \cellcolor[HTML]{FFD966}0.354 & \cellcolor[HTML]{FFD966}0.002 & \cellcolor[HTML]{FFD966}0.017 & \cellcolor[HTML]{FFF9E7}0.001   \\ \midrule
                                & DP                             & 0.961                         & 0.974                        & 0.949                        & 0.947                           & 0.948                           & 0.968                                      & \cellcolor[HTML]{FFD966}0.982  & \cellcolor[HTML]{FFD966}0.970     & \cellcolor[HTML]{FFD966}0.011      & \cellcolor[HTML]{FFFFFF}0.000 & \cellcolor[HTML]{FFFFFF}0.000 & \cellcolor[HTML]{FFFFFF}0.000 & \cellcolor[HTML]{FFFFFF}0.000 & \cellcolor[HTML]{FFFFFF}0.000 & \cellcolor[HTML]{FFFFFF}0.000 & \cellcolor[HTML]{FFFEFB}0.000   \\
                                & HMM                            & 0.948                         & 0.982                        & 0.925                        & 0.886                           & 0.925                           & 0.901                                      & \cellcolor[HTML]{FFE38B}0.977  & \cellcolor[HTML]{FFEAAA}0.914     & \cellcolor[HTML]{FFFEFA}0.002      & \cellcolor[HTML]{FFEAA8}0.005 & \cellcolor[HTML]{FFFFFF}0.000 & \cellcolor[HTML]{FFFFFF}0.000 & \cellcolor[HTML]{FFF2C8}0.047 & \cellcolor[HTML]{FFF0BF}0.010 & \cellcolor[HTML]{FFDF7B}0.000 & \cellcolor[HTML]{FFFFFC}0.000   \\
\multirow{-3}{*}{Rock}          & SPPK                           & 0.915                         & 0.972                        & 0.879                        & 0.771                           & 0.827                           & 0.820                                      & \cellcolor[HTML]{FFFFFF}0.962  & \cellcolor[HTML]{FFFFFF}0.843     & \cellcolor[HTML]{FFFFFF}0.001      & \cellcolor[HTML]{FFD966}0.009 & \cellcolor[HTML]{FFD966}0.002 & \cellcolor[HTML]{FFD966}0.001 & \cellcolor[HTML]{FFD966}0.129 & \cellcolor[HTML]{FFD966}0.025 & \cellcolor[HTML]{FFD966}0.000 & \cellcolor[HTML]{FFEDB7}0.003 

\\\bottomrule
\end{tabular}
\label{tab:allmet_res}
\end{table*}


As for the beat trackers, we adopt the two-block architecture mentioned in Section \ref{sec:intro} and implement three approaches.
They all 
use the popular open-source library,  \texttt{madmom}~\cite{Bock2016mm, Bock2016d} to generate the beat activation functions $\Delta(n)$,
but  use different PPTs for converting $\Delta(n)$ into the estimated beat positions $\hat{B}$.
Specifically, we use
a simple peak picking (SPPK) method \cite{scipy}, a DP-based tracker \cite{DP2007,fmp2021, libfmp2021}, and an HMM-based tracker \cite{Krebs2015, Bock2016mm}. 
For SPPK, we use the ``find\_peaks'' function of the \texttt{scipy} library \cite{scipy}.
For HMM, we use the implementation in \texttt{madmom}, using the default setting. 
The DP  requires a global tempo value to be provided beforehand. Following \cite{Grosche2011}, we compute the global tempo from the mean IBI of the reference beats. 
Note that no training is involved in our experiment because the network provided by \texttt{madmom} is already trained and the three PPTs do not need any training.


%

\section{Experiments}


Table \ref{tab:allmet_res} shows the evaluation results of the beat trackers using the proposed ACR and existing metrics including  F1-score, CMLt/AMLt, and L-correct ($\mathrm{F}^{L}$).
According to the result of the existing metrics, we can see that beat tracking seems to be easier for music with steady tempo (cf. Table \ref{tab:dataset}): for example, all the three trackers have F1 scores above 0.900 on the Rock dataset.
However, a closer look reveals that there is inconsistency among the results of the existing metrics.
We make the following observations.
{\textbf{O1}}: 
For both Maz-5 and Jazz, HMM slightly outperforms DP in F1, but, \cyc{without explanation}, DP gets much better result in L-correct and CMLt/AMLt metrics.
{\textbf{O2}}: 
From the low \sunny{$\mathrm{F}^{L}$} (0.128), low recall  (0.393), and high precision  (0.752) of HMM for Maz-5, we conjecture that HMM may tend to tap at slower tempi than the reference beats (therefore cannot match the reference beats consecutively for $L=2$), but none of the existing metrics clearly support this. 
{\textbf{O3}}: Similarly, 
compared to DP, HMM has lower L-correct, higher recall and lower precision for Jazz, suggesting that HMM may 
tap at faster tempi here, 
but again more evidence is needed.


Our newly proposed ACR metrics provide explanations to these observations
and lead to new insights.
For example, from the ACR scores for Maz-5 in Table \ref{tab:allmet_res}, one can understand that HMM not only taps at slower tempi but also switches among different tempo subharmonics, which explains its extremely low scores \sunny{w.r.t.} L-correct and CMLt/AMLt metrics. 
Similarly, the ACR scores show that HMM does tap faster and mainly at double tempo for Jazz. 
It seems that HMM tends to tap slower at subharmonics for faster tracks (e.g., Maz-5) and tap faster at harmonics for slower tracks (e.g., Jazz), \sunny{explaining} its inferior scores in existing metrics such as L-correct compared to DP.
Besides, from the ACR of DP, we see that although DP is exempt from the harmonic/subharmonic ``errors,'' it instead suffers from offbeat ``errors,'' as also reported in \cite{Grosche2011}.
\sunny{In sum, for challenging music pieces, existing beat trackers have not learned to adapt to the tempo changes, and could produce different types of metric-level \cyc{or phase-}related ``errors’’ that lead to poor scores not fully explainable by existing metrics. The ACR reveals these metric-level behaviors and provides additional evaluation perspectives. For example, according to ACR-any, if MLS is allowed in the middle of a music piece, HMM is actually not inferior to DP. 
}



\begin{table}[t]
\caption{Context sensitive evaluation result with $L=2, 3, 4$.}
\setlength{\tabcolsep}{5pt}
\centering
\begin{tabular}{l|c|ccc|ccc}
\toprule
\multirow{2}{*}{\textbf{Data}} & \multirow{2}{*}{\textbf{PPT}} & 
\multicolumn{3}{c|}{\textbf{ACR-onbeat}} & \multicolumn{3}{c}{\textbf{ACR-any}}  \\\cline{3-8}
%
%
& & \textbf{L=2} & \textbf{L=3} & \textbf{L=4} & \textbf{L=2} & \textbf{L=3} & \textbf{L=4} \\\midrule
\multirow{3}{*}{Maz-5}         & DP                            & 0.383        & 0.310        & 0.243        & 0.598        & 0.414        & 0.294        \\
                               & HMM                           & 0.112        & 0.102        & 0.096        & 0.726        & 0.597        & 0.484        \\
                               & SPPK                          & 0.682        & 0.622        & 0.566        & 0.904        & 0.721        & 0.622        \\ \midrule
\multirow{3}{*}{\cyc{Jazz}}      & DP                            & 0.790        & 0.787        & 0.784        & 0.915        & 0.904        & 0.895        \\
                               & HMM                           & 0.669        & 0.667        & 0.666        & 0.903        & 0.893        & 0.887        \\
                               & SPPK                          & 0.452        & 0.366        & 0.324        & 0.774        & 0.658        & 0.576        \\\midrule
\multirow{3}{*}{Rock}          & DP                            & 0.970        & 0.969        & 0.969        & 0.982        & 0.979        & 0.977        \\
                               & HMM                           & 0.914        & 0.914        & 0.914        & 0.977        & 0.977        & 0.976        \\
                               & SPPK                          & 0.843        & 0.813        & 0.797        & 0.962        & 0.931        & 0.906  \\ \bottomrule     
\end{tabular}
\label{tab:moreL}
\end{table}

Table \ref{tab:moreL} further shows the ACR results for the metrical level of onbeat and `any tempo' with $L=2, 3, 4$, which
provides insights for how well a beat tracker can handle the temporal context of several consecutive beats, rather than looking at the beats individually. For Maz-5, using either ACR-onbeat or ACR-any, 
the scores for the three PPTs all drop remarkably as $L$ increases, indicating that none of them can handle well the temporal context. On the contrary, for Jazz and Rock, the scores of DP and HMM remain almost unchanged as $L$ increases, indicating that most of the estimated beats are predicted \sunny{correctly for a sequence of concecutive beats} rather than individually. From the remarkably higher ACR-any values compared to ACR-onbeat values of DP and HMM for Jazz, we also see that they can handle 
\sunny{more than $90\%$ of}
the temporal context well (i.e., ACR-any $>0.90$)
but with metric-level switching.  
From the 
performance degradation of SPPK as $L$ increases, we see that SPPK, which makes estimation merely based on individual activation peaks, cannot handle temporal context as HMM and DP can do \cyc{for \cyc{Jazz} and Rock}. 

\section{Conclusion}
In this paper, we introduced an analysis method, called annotation coverage ratio (ACR), to reveal the metric-level switching behaviors of beat trackers. With experiments on datasets of three major music genres, we demonstrate how the proposed ACR can compensate for the inadequacy of existing evaluation metrics and facilitate further analysis of beat tracking for challenging genres of music (e.g., classical and jazz). We hope this work can contribute to the development of more advanced and general beat trackers.

\bibliographystyle{IEEEtran}
\bibliography{acr_spl}

\end{document}